\documentclass[%
 reprint,
 amsmath,amssymb,
 aps,
]{revtex4-2}

\usepackage{graphicx}
\usepackage{dcolumn}
\usepackage{bm}
\usepackage{xcolor}

\begin{document}

\preprint{APS/123-QED}

\title{Numerical investigation of viscous fingering in a three-dimensional cubical domain}

\author{Garima Varshney}
\author{Anikesh Pal}%
 \email{pala@iitk.ac.in}
\affiliation{%
 Department of Mechanical Engineering, Indian Institute of Technology Kanpur, 208016, India.\\
}%




\date{\today}

\begin{abstract}

We perform three-dimensional numerical simulations to understand the role of viscous fingering in sweeping a high-viscous fluid (HVF). These fingers form due to the injection of a low-viscous fluid (LVF) into a porous media containing the high-viscous fluid. We find that the sweeping of HVF depends on different parameters such as the Reynolds number ($Re$) based on the inflow rate of the LVF, the P\'eclet number ($Pe$), and the logarithmic viscosity ratio of HVF and LVF, $\mathfrak{R}$. At high values of $Re$, $Pe$, and $\mathfrak{R}$, the fingers grow non-linearly, resulting in earlier tip splitting of the fingers and breakthrough, further leading to poor sweeping of the HVF. In contrast, the fingers evolve uniformly at low values of $Re$, $Pe$, and $\mathfrak{R}$, resulting in an efficient sweeping of the HVF. We also estimate the sweep efficiency and conclude that the parameters $Re$, $Pe$ and $\mathfrak{R}$ be chosen optimally to minimize the non-linear growth of the fingers to achieve an efficient sweeping of the HVF.\\

\end{abstract}

\maketitle


\section{\label{sec:level1}Introduction 
}
Finger-like protrusions \citep{hill1952channeling, saffman1958penetration} form when a low viscous fluid (LVF) displaces a high viscous fluid (HVF) in a porous medium owing to hydrodynamic instability along the interface of these two fluids. This type of hydrodynamic instability of multi-phase flow is often referred to as viscous fingering and appear in many engineering and scientific processes such as oil recovery from underground reservoirs \citep{joekar2012analysis,muggeridge2014recovery}, chromatography \citep{shalliker2007visualising}, $CO_2$ sequestration \citep{berg2012stability,mohanty2015development}, fluid mixing in microfluidics \citep{jha2013synergetic}, and oceanography \citep{coumou2006dynamics,pal2020evolution}. The association of the viscous fingering phenomenon in multiple areas has motivated researchers to investigate its dynamics theoretically \citep{hill1952channeling,tan1987stability,waggoner1990growth,shelley1997hele,derks2003cohesive,dias2010control}, experimentally \citep{paterson1981radial,bischofberger2014fingering,rabbani2018suppressing} and numerically \citep{bogdanov2007comsol,pramanik2012miscible,wijeratne2015computational,sharma2016fingering, vishnudas2017comprehensive, garaeva2019numerical}.\\

The pioneering work on viscous fingering encountered during sugar refining operations, was carried out by \cite{hill1952channeling}. They referred to these instabilities as "channelling" when water displaces sugar liquors from columns of granular bone charcoal. The next significant development, that occurred in the late 1950s, established \cite{saffman1958penetration,chuoke1959instability} that adverse mobility (when an LVF displaces an HVF) generates an unstable interface. A review of the experiments and the numerical simulations performed to study the mechanisms of viscous fingering in homogeneous porous materials using different physical models and geometries (rectilinear displacement, radial source flow, and the five-spot pattern) is provided in \cite{homsy1987viscous}. A description of the development of the Saffman-Taylor instability in a two-dimensional (2D) porous media (also represented by a Hele-Shaw cell due to the opaque nature of porous media) owing to the convection-diffusion phenomenon for miscible fluid interaction is also provided in this review. Moreover, it was deduced that the P\'eclet number was one of the primary parameters that govern the fingering scale for miscible fluids. Viscous fingering in the miscible and immiscible fluids was also experimentally investigated in a Hele-Shaw cell with smooth and etched plates to study the influence of plate roughness on the fingering mechanism by \cite{chen1987radial}. They reported that owing to interfacial tension, the immiscible finger patterns are less ramified than their miscible counterparts, are more sensitive to the flow rate, and become compact as the flow rate decreases. Experiments in a real three-dimensional (3D) porous medium were performed by \cite{bacri1991three} to study the essential features of viscous fingering and its dependence on the viscosity ratio and the flow rate. It was observed that the single-finger configuration was strongly affected by adding a small perturbation in the gap of the Hele-Shaw cell \citep{zhao1992perturbing}. \cite{loggia1999effect} performed experiments on the miscible displacements in porous media to study the effects of mobility gradients in viscous instability and corroborated their findings with an analytical analysis. A rectilinear Hele-Shaw cell was also used to examine the miscible flow displacements of a reference Newtonian fluid (glycerol solution) or shear-thinning solutions of Alcoflood polymers of different molecular weights by water through experimental measurements \citep{li2006experimental}. Similarly, a radial Hele-Shaw cell was used \citep{bischofberger2014fingering} to explore the variety of patterns characterized by the viscosity ratio of the two interacting fluids. Recently, an experimental study \citep{kargozarfard2019viscous} was carried out using the Hele-Shaw cell to explore the techniques to suppress the viscous fingering problem encountered.\\

Several numerical studies using the Hele-Shaw cell model, either in radial or rectilinear shape, have been conducted in addition to the experimental work. Steam-assisted gravity drainage, a thermal oil recovery method, is explored \citep{bogdanov2007comsol} by integrating mass balance and energy balance using the commercial solver COMSOL for a two-dimensional domain and comparing the outcomes with those from another reservoir simulator, STARS. Subsequently, COMSOL's two-phase Darcy's law physics is employed in a 2D Eulerian frame to study the instability in chromatographic columns and aquifers for various injection speeds and mobility \citep{pramanik2012miscible}. A similar setup was also used to simulate the suppression, decrease, or increase of miscible viscous fingering in radial displacements in a 2D homogeneous porous medium by varying the mobility ratio, injection speed, and diffusion coefficient \citep{sharma2016fingering}. The commercial solver ANSYS has been used to perform 2D simulations \citep{wijeratne2015computational} to study the viscous fingering phenomenon associated with oil production in a homogeneous heavy oil reservoir. An open-source solver UTCHEM was used by \cite{vishnudas2017comprehensive} to perform 2D simulations of miscible and immiscible viscous fingering encountered during polymer flooding chemical enhanced oil recovery processes. To capture the dynamic evolution of these viscous fingers, they performed a Fourier analysis of the saturation or the concentration contours and the rate of change of root-mean square (RMS) of the saturation/concentration contours. Recently, a 2D simulation was carried out by \cite{garaeva2019numerical} to investigate the effect of the period and amplitude of the initial boundary perturbation on the growth rate of viscous fingers. A comprehensive review \citep{pinilla2021experimental} is provided for the experimental and the computational investigations of miscible and immiscible fluid interaction.\\

As evident from the previous discussion, majority of the numerical studies performed to understand the dynamics of viscous fingering are two-dimensional. The first 3D simulation of miscible viscous fingering is reported by \cite{zimmerman1992three} at a high P\'eclet number. It was concluded that the mechanism of nonlinear interactions of viscous fingers in 3D is in accord with the observations made from 2D simulations. Similar to \cite{zimmerman1992three}, \cite{tchelepi1993dispersion} also reported that the essential features of fingering in a homogeneous porous medium or a porous medium with modest, randomly distributed heterogeneities obtained from a 3D simulation could also be represented by a 2D calculation. However, permeability variations and connectivity in the third dimension may influence the fluid distribution in flow through heterogeneous domains with significantly correlated scales. Therefore, 3D modeling will be imperative for reproducing experimental results. Moreover, 3D modeling will also be inevitable if there is a density difference between the displacing and resident fluid in a homogeneous porous medium and the mean flow direction is not vertical. The impact and importance of 3D effects in viscous fingering for increasing density difference between the displacing and resident fluid in water alternating gas (WAG) injection in oil recovery are also reported \cite{christie19933d}. They concluded that for a range of WAG ratios the 3D computation exhibits a lower recovery and an earlier oil breakthrough than 2D simulations. Another numerical simulation was performed by \cite{riaz2003three} to analyze the three-dimensional miscible displacements with gravity override in a homogeneous porous medium in the quarter five-spot geometry. They demonstrated that the enhanced interaction of the disturbances in 3D alters the character of the flow in a manner that could not be captured by 2D simulations. The present investigation examines the evolution of viscous fingering, owing to the interaction of low-viscosity and high-viscosity miscible fluids in a three-dimensional homogeneous porous medium, under the influence of different parameters such as the flow rate of displacing fluid, different encapsulated fluid with variation in the viscosity, and ease of diffusion through the porous domain. The problem formulation, governing equations, numerical methodology, and case set-up are presented in section \ref{sec:problem}. The results obtained from the numerical simulations for the various parameters are discussed in section \ref{sec:results}, and conclusions are drawn in section \ref{sec:conclusion}.\\

\section{\label{sec:problem}Problem formulation} 

We define the fundamental parameters to understand the characteristics of 3D fingering instability and their non-linear interaction \citep{waggoner1990growth,pinilla2021experimental}. The logarithmic mobility ratio ($\mathfrak{R}$) is the logarithmic ratio between the viscosity of the displaced fluid and the displacing fluid, $\mathfrak{R} = ln\left (\frac{\mu_2}{\mu_1}\right)$; porosity ($\epsilon_p)$ is the storing capacity of fluid in the porous material, which represents the ratio of the volume occupied by the fluid, and the total volume of the porous material; breakthrough defines the moment when LVF reaches the downstream of the domain; the sweep efficiency is ($\eta_{sw}$) defined as the ratio of the volume of the LVF injected at the breakthrough time to the volume of the domain; shielding is the tendency for one finger to dominate the displacement due to the finite amount to the injected fluid; spreading is the widening and flattening of the tip and body of the figures as it grows;  tip splitting is the creation of two small fingers at the tip due to the instability of the tip of a larger finger \citep{waggoner1990growth}; coalescence is the merging of the tip of one finger into the body of an adjacent finger \citep{li2006experimental}; P\'eclet number ($Pe$): is defined as the ratio of the rate of advection to diffusion rate of a fluid $Pe= \frac{\vec U L_c} {\mathfrak{D}} $ where $\vec U, L_c,$ and $\mathfrak{D}$ are the average velocity of flow, characteristic length (the diameter of hole through which LVF is injected into the domain is taken as the characteristics length in this study), and diffusion coefficient; Reynolds number ($Re$) is the ratio of inertia forces to the viscous forces, defined as $ Re= \frac{\rho \vec U L_c}{\mu}$ , volume flow rate ($\dot{Q}$), and permeability ($ \kappa $) regulates the capacity of the fluids to flow through porous media. \\

\begin{figure}  
\includegraphics[width=0.99\linewidth]{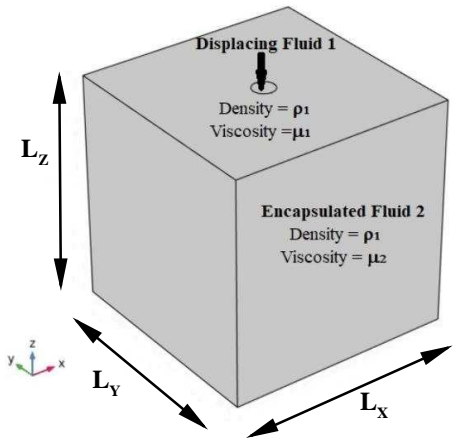}
\caption{\label{fig:1} Schematic of a cubical porous computational domain having an encapsulated fluid 2 (HVF) and a cylindrical hole to inject fluid 1 (LVF).}
\end{figure}

\begin{figure}  
\includegraphics[width=0.99\linewidth]{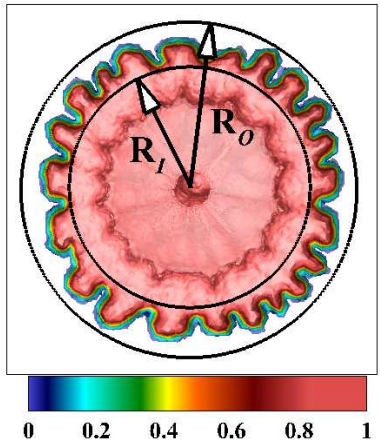}
\caption{\label{fig:2}Definition of different length scales $R_I, R_O$ for quantifying finger growth.}
\end{figure}


\subsection{\label{sec:level2}Governing Equations and Numerical Method }

COMSOL Multiphysics 6.0 \citep{COMSOL2021} is used to model the viscous fingering phenomenon. We explore the characteristics of these fingers by varying the mass flow rate of the displacing fluid, the diffusion coefficient, and the mobility of the fluid encapsulated in the domain. The associated non-dimensional parameters are the Reynolds number($Re$), the P\'eclet number($Pe$), and the log mobility ratio($\mathfrak{R}$). The two-phase Darcy's law model of porous media and subsurface fluid flow module is used, which couples the steady Darcy flow equation with the time-dependent convection-diffusion equation for concentration. The governing equations associated with this module are:\\
 
\begin{eqnarray}
\frac{\partial \varepsilon_p \rho}{\partial t} + \nabla \cdot \rho \vec{u} =0
\label{eq:one}.
\end{eqnarray}

\begin{eqnarray}
 \vec{u} = - \frac{\kappa}{\mu} \vec{\nabla}p
\label{eq:two}.
\end{eqnarray}

\begin{eqnarray}
   \rho = s_1 \rho_1 +s_2 \rho_2
\label{eq:three}.
\end{eqnarray}

\begin{eqnarray}
   \frac{1}{\mu}= s_1 \frac{\kappa_{r1}}{\mu_1}+s_2 \frac{\kappa_{r2}}{\mu_2}
\label{eq:four}.
\end{eqnarray}

 \begin{eqnarray}
   s_1 +s_2=1
\label{eq:five}.
\end{eqnarray}

\begin{eqnarray}
  \frac{\partial \varepsilon_p c_1}{\partial t} + \vec{\nabla}\cdot (c_1\vec{u})= \Vec{\nabla}\cdot (\mathfrak{D}_c  \vec{\nabla}c_1)
\label{eq:six}.
\end{eqnarray}
   
\begin{eqnarray}
  c_1=s_1\cdot\rho_1 
\label{eq:seven}.
\end{eqnarray}

\begin{eqnarray}
 \mu(s) = \mu_2\cdot e^{-\mathfrak{R}\cdot s_1}
\label{eq:eight}.
\end{eqnarray}

\begin{eqnarray}
 s_1= 0.5(1+\zeta(f(x,y,z)))
\label{eq:nine}.
\end{eqnarray}

Here suffixes $1$ and $2$ denote the characteristics of fluids 1 (LVF) and 2 (HVF). We consider two fluids of different saturation values, $s_1$ and $s_2$, related by equation (\ref{eq:five}). Additionally, the saturation value $s_1$ and the concentration $c_1$ are related via equation (\ref{eq:seven}). This module solves the Darcy's law for total pressure and the convection-diffusion equation for fluid transport. We discretize the governing equations for both variables velocity and pressure using the finite element method with second-order quadratic Lagrange elements  \citep{bogdanov2007comsol, pramanik2012miscible, holzbecher2009modeling}. We have used the implicit backward differentiation formula (BDF), with the maximum and minimum degrees of the interpolating polynomial being $5$ and $1$, respectively. The time-step is controlled nonlinearly owing to its effectiveness over the free-step sizing. We use the fully coupled, constant Newton technique to solve nonlinear systems, and an iterative solver is used to solve the linear systems. We use successive over-relaxation (SOR) as a pre-and post-smoother, and parallel direct solver (PARDISO) is used as a coarser solver when the generalized minimal residual method (GMRES) is employed in the multigrid solver.\\

The computational domain used for this investigation is a 3D cubical homogeneous porous region of porosity $\epsilon_p$ and permeability '$ \kappa $' as shown in figure \ref{fig:1}. We inject LVF of viscosity, $\mu_1$, with a vertical velocity of $u = U_0$. This LVF sweeps a HVF of viscosity, $\mu_2 $, encapsulated in the porous domain. The cylindrical inlet from where the injection of LVF occurs has a diameter and height of $ 0.25$ mm and $0.06$ mm, respectively. The size of the cubical domain is $2.5 mm {^3}$. To avoid the fingering caused by density variations, we assume that both fluids are of the same density, Newtonian in nature, incompressible, and do not react with each other. Therefore, the dynamics of these miscible fluids will primarily be affected by the concentration and the pressure gradients, while mobility differences will govern the dynamics of the viscous fingering. We use tetrahedral mesh with a minimum element size of 0.5 $\mu$m and approximately three million total elements across the domain. The inflow velocity and concentration are $U_0$ and $s_1=1$. For the pressure at inlet, we use the homogeneous Neumann boundary condition. At the outlet, homogeneous Neumann boundary conditions are used for the velocity, whereas, for the pressure, we use homogeneous Dirichlet boundary condition ($p=0$). Also, to ensure no flux at other transverse boundaries Neumann boundary condition, $-n\cdot \rho u = 0$, is employed. The initial values of $s_1 = 0$ in the cubical domain. A linear perturbation profile in the form of an arbitrary three-dimensional function $f$ (x, y, z) and amplified using $\zeta$ (equation \ref{eq:nine}) is imposed at the junction of the cubical domain and cylindrical inlet. This setup is relevant to the oil industry, where a fluid (LVF) is injected to sweep another fluid (HVF) trapped in a porous domain. The viscosity mismatch between LVF and HVF results in viscous fingering, and the flow will be unstable. We define the length scales as shown in figure \ref{fig:2} \citep{bischofberger2014fingering} to quantify the fingering patterns. The largest circle that encloses the region where injected fluid sweeps out the encapsulated fluid is known as the inner radius, $R_I$. The outer radius, $R_O$, is the smallest circle that encompasses the most distantly displaced fluid, and $R_F= R_O-R_I$ represents the maximum expansion of the fingers at that time. The list of parameters used in the study is given in table \ref{tab:table1}.\\

\begin{table*}
\caption{\label{tab:table1}List of parameters}
\begin{ruledtabular}
\begin{tabular}{ccc}
 $\textbf{Parameters}$ & \textbf{Detail of Parameters} &\textbf{Values}
 \\ \hline
 $L_x=L_y=L_z$ & Length, Width, and Height of cubical domain & $ 2.5 mm, 5mm$ \\
r and h & Radius and Height of Injected hole
 &0.125 mm, 0.06 mm\\
 $\mathfrak{R}$ & Logarithmic mobility ratio & 2,3,4 \\
$U_o$ & Injection speed & 0.5, 1, 2 m/s \\
$\mu_1$ & Viscosity of displacing fluid & 0.001Pa-s \\
$\rho$ & Density of both the fluids  & $1.225 Kg/m^3 $\\
$\epsilon_p$ & Porosity  & 0.5 \\
$\kappa$ & Permeability  & $10^{-6} m^2$ \\
$\mathfrak{D}$ &Diffusion coefficient& $ 2\times10^{-8},2\times10^{-9},2\times10^{-7} m^2/s $\\
$\mathfrak{D}_c$  & Capillary Diffusion coefficient &  $\epsilon_p\times \mathfrak{D} $\\
$\zeta$ & Amplitude of the disturbance & 0.01 \\
$\kappa_{r1}=\kappa_{r2}$ &Relative Permeability of Fluids  & 1\\
$Re, 2Re, 0.5Re$ & Reynolds Number & 0.30625,0.6125,0.153125 \\
$Pe, 10Pe, 0.1Pe$ & P\'eclet Number & $ 1.25\times10^4, 1.25\times10^5, 1.25\times10^3 $ \\

\end{tabular}
\end{ruledtabular}
\end{table*}

\begin{table*}
\caption{\label{tab:table2} Cases for the study}
\begin{ruledtabular}
\begin{tabular}{ccccc}

\textbf {Cases} & $\textbf{Re}$ & $$\textbf{Pe}$$ & $$\textbf{$\mathfrak{R}$}$$ &\textbf{Nomenclature}\\\hline
Case 1 & $Re$=0.30625 & $Pe$=1.25$\times10^4$&2& C1($Re,Pe$,2) \\
Case 2 & $0.5Re$=0.153125 & $Pe$=1.25$\times10^4$&2 & C2($0.5Re,Pe$,2) \\
Case 3 & 2$Re$=0.6125 & $Pe$=1.25$\times10^4$& 2 & C3($2Re,Pe$,2)\\
Case 4 & $Re$=0.30625 & 0.1$Pe$=1.25$\times10^3$ &2 & C4($Re,0.1Pe$,2)\\
Case 5 & $Re$=0.30625  & $10Pe$=$1.25\times10^5$&2 & C5($Re,10Pe$,2)\\
Case 6 & $Re$=0.30625 & $Pe$=1.25$\times10^4$&3 & C6($Re,Pe$,3)\\
Case 6* (Double sized domain) & $Re$=0.30625 & $Pe$=1.25$\times10^4$&3 & C6$^{*}$($Re,Pe$,3)\\
Case 7 & $Re$=0.30625 & $Pe$=1.25$\times10^4$&4 & C7($Re,Pe$,4)\\

\end{tabular}
\end{ruledtabular}
\end{table*}

\section{\label{sec:results}Results and discussion}

\begin{figure*}
(a) {\includegraphics[width=0.9\linewidth]{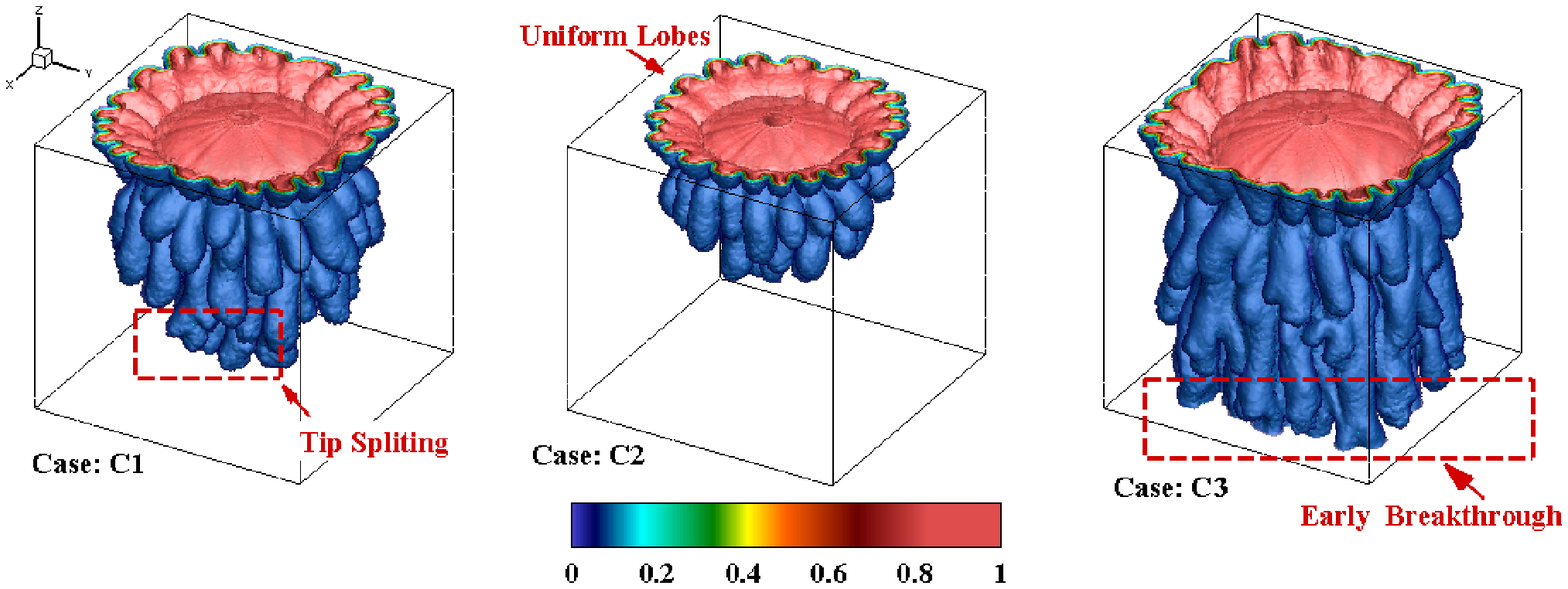}}\\
(b) {\includegraphics[width=0.9\linewidth]{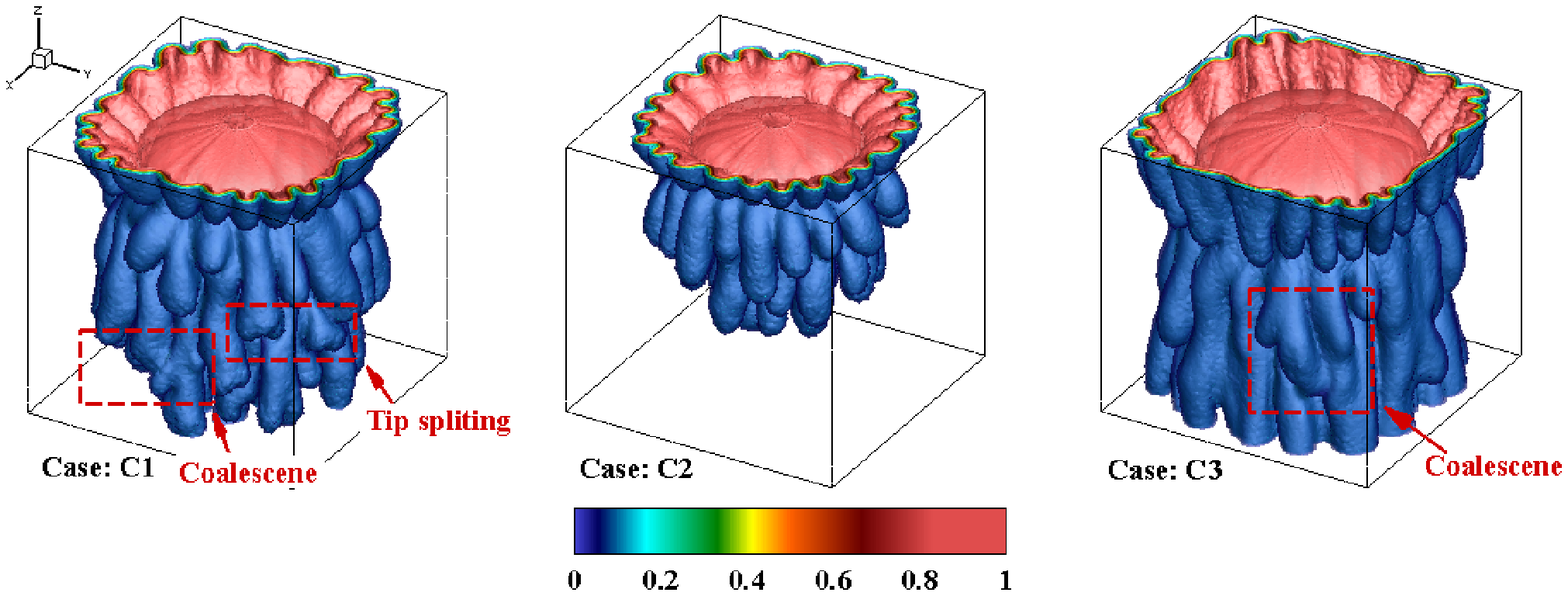}}
\caption{\label{fig:3}Iso-contours of LVF saturation ($s_1$) at times (\textit{a}) $40$ ms, and (\textit{b}) $60$ ms for the cases with different $Re$.}
\end{figure*}

Hydrodynamic instability in energy resource production reduces sweep efficiency. This results in decreased oil production. Permeability differences between the fluids, viscous forces, gravity forces, capillary forces, and diffusion caused by concentration differences are the key factors that control such complicated phenomena. To study the role of such regulating factors for the instabilities in more diversified ways, we simulate cases for miscible fluids with different non-dimensional numbers as indicated in table \ref{tab:table2}. A comparison of the saturation iso-surfaces at different LVF injection velocities while keeping all other factors the same is shown in figures \ref{fig:3}(a) and (b) at $40$ ms and $60$ ms, respectively. For case C2, the fingers are approximately uniform in shape and size. However, with increasing $Re$ (cases C1 and C3), the fingers become wider and grow non-uniformly. Early breakthroughs will occur for higher injection velocity, along with early tip splitting. The fingers eventually converge due to the coalescing phenomenon when the instabilities approach the domain walls. The structure of the fingers has a substantial variation for different flow rates at the same time instances, as observed in figures \ref{fig:3}(a) and (b). \\

\begin{figure*}
(a) {\includegraphics[width=0.9\linewidth]{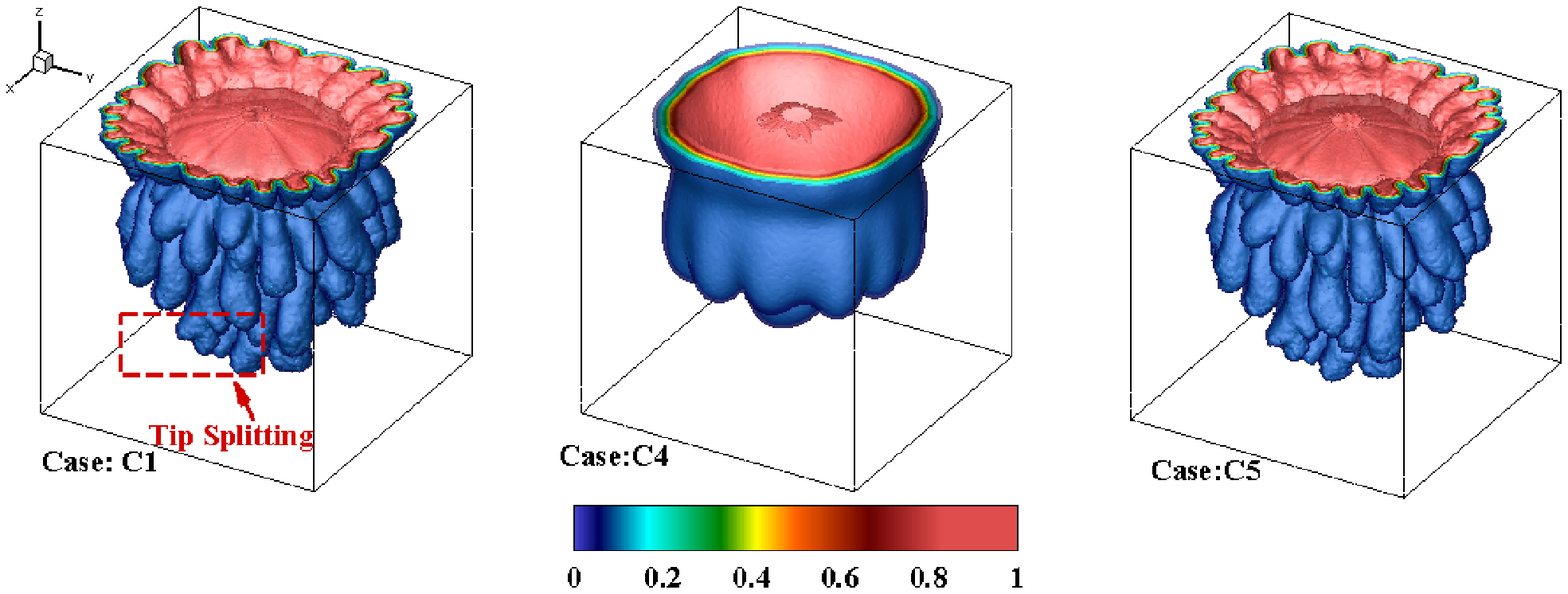}}\\
(b) {\includegraphics[width=0.9\linewidth]{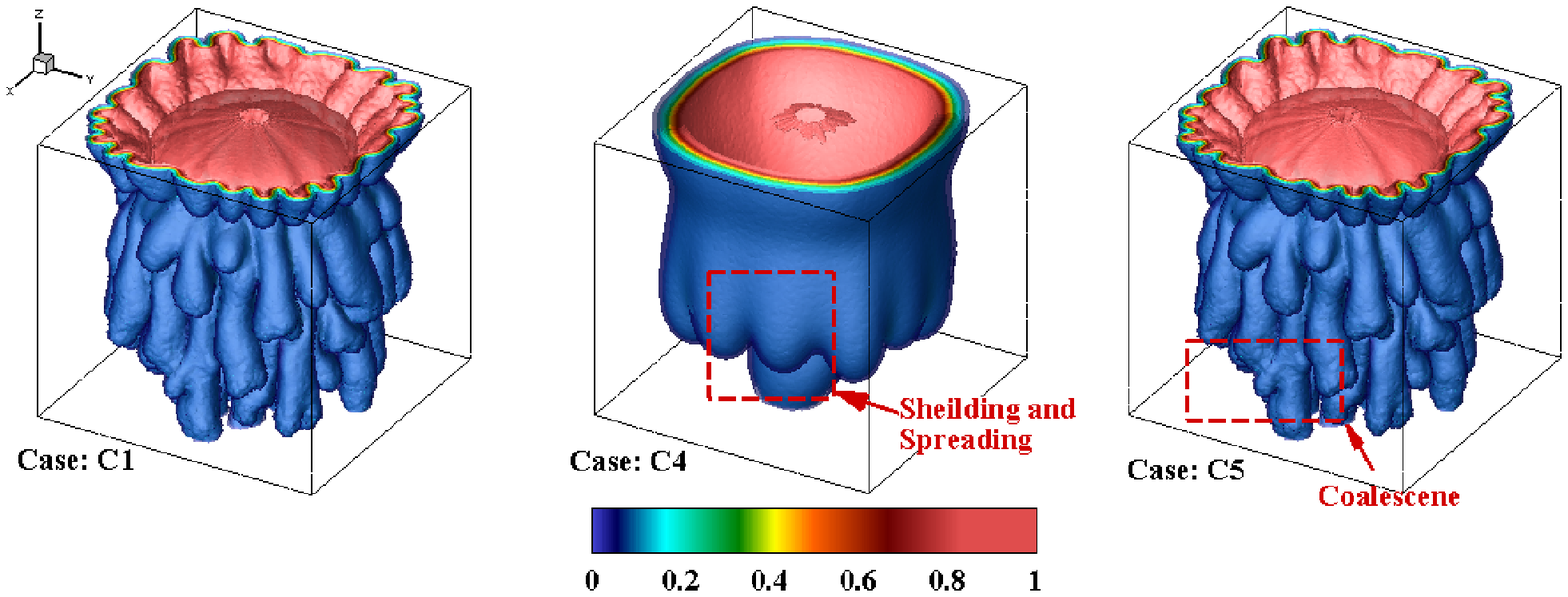}}
\caption{\label{fig:4} Iso-contours of LVF saturation ($s_1$) at times (\textit{a}) $40$ ms, and  (\textit{b}) $60$ for cases C1, C4, and C5 having variation in $Pe$ as $Pe, 0.1Pe, 10Pe$ respectively.}
\end{figure*}

The evolution of the flow occurs due to pressure difference and diffusion, governed by the Darcy and the convection-diffusion equation, respectively. The relationship between temperature, pressure, and the properties of the diffusing substance, such as the size of its molecules, affect the diffusion coefficient. To investigate the impact of changing the diffusion coefficient and keeping all the other parameters the same, we simulate cases by varying the values of $Pe$ (see table \ref{tab:table1}). Figures \ref{fig:4}(a) and (b) illustrate the iso-surface for LVF's saturation for the cases with variation in $Pe$ (Case C1, C4, and C5) at $40$ and $60$ ms, respectively. When $Pe$ is low (case C4), the diffusion of LVF dominates the viscous effects, and the fingering instabilities cease to evolve. The absence of finger-like patterns at the cross-sections of the domain signifies that the effect of the mobility difference is negligible for C4. The central finger displaces the encapsulated HVF because of shielding and spreading at low $Pe$ (Case 4). Eventually, this central finger reaches downstream. For this case, our results are qualitatively comparable to the results of \citep{suekane2017three} for the same order of P\'eclet number but a higher viscosity ratio.\\

\begin{figure*}
{\includegraphics[width=0.9\linewidth]{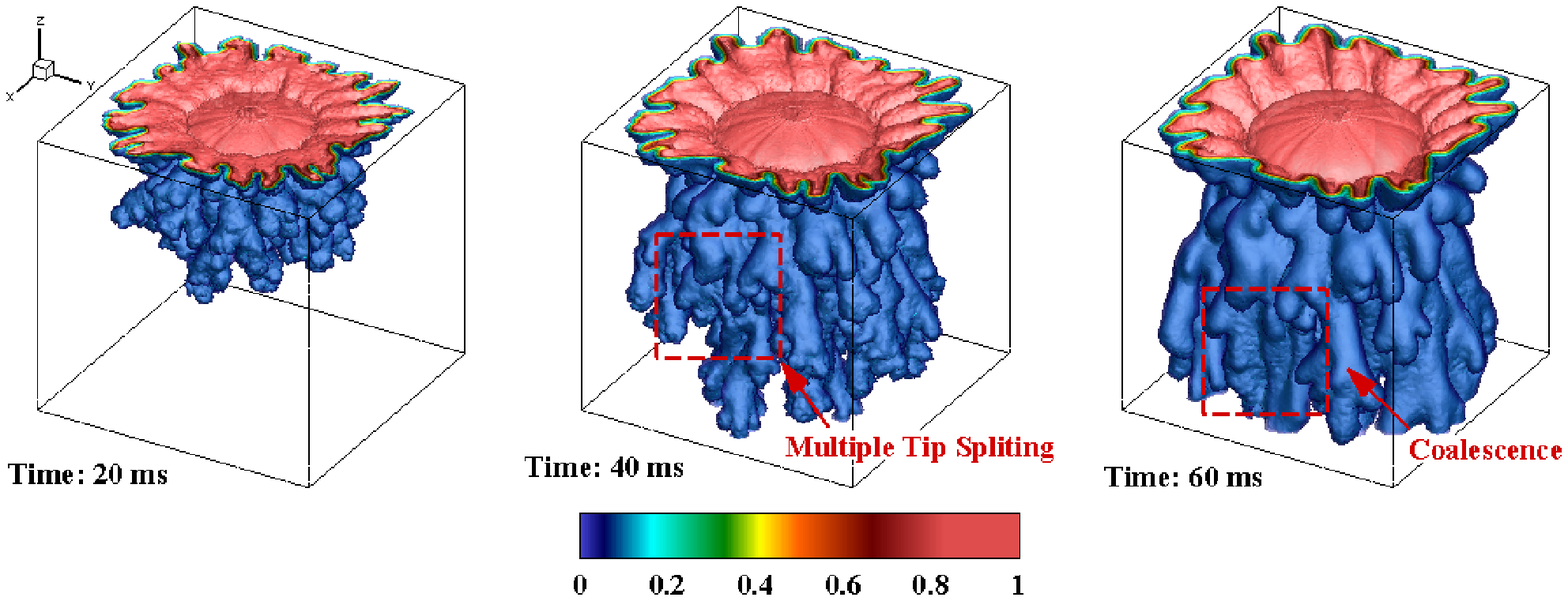}}
\caption{\label{fig:5} Iso-contours of LVF saturation ($s_1$) for case C6($Re,Pe$,3) at times $20$, $40$, and $60$ ms respectively.}
\end{figure*}

\begin{figure*}
{\includegraphics[width=0.9\linewidth]{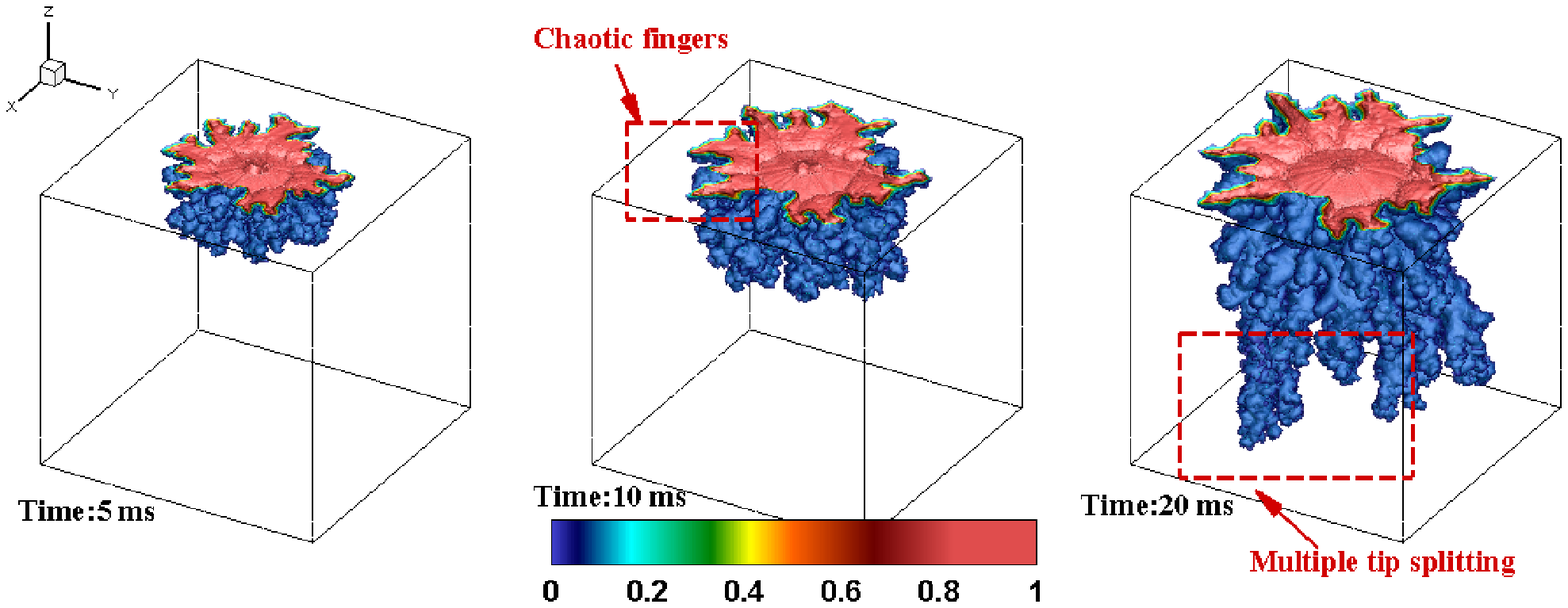}}
\caption{\label{fig:6} Iso-contours of LVF saturation ($s_1$) for case C7($Re,Pe$,4) at times $5$, $10$, and $20$ ms respectively.}
\end{figure*}

Figure \ref{fig:5} compares the evolution of the instability for case C6 for the logarithmic viscosity ratio $\mathfrak{R}= 3$ at times $20$, $40$, and $60$ ms while figure \ref{fig:6} compares the patterns of instability for case C7 ($\mathfrak{R} = 4$) at times $5$, $10$ and $20$ ms. With an increase in $\mathfrak{R}$, tip splitting of the fingers occur, which results in thinner fingers that travel downstream. For C1 (where $\mathfrak{R} = 2$), the fluid expands more uniformly and sweeps the majority of the encapsulated fluid, whereas in C6 and C7 ($\mathfrak{R} = 3,4$ respectively), owing to the disordered finger growth lesser amount of fluid is swept out, before the breakthrough. For larger values of $\mathfrak{R}$, the chaotic and nonlinear growth of the fingers occurs with earlier tip splitting, making the sweeping of the HVF by the LVF more challenging.\\

\begin{figure}  
\includegraphics[width=0.99\linewidth]{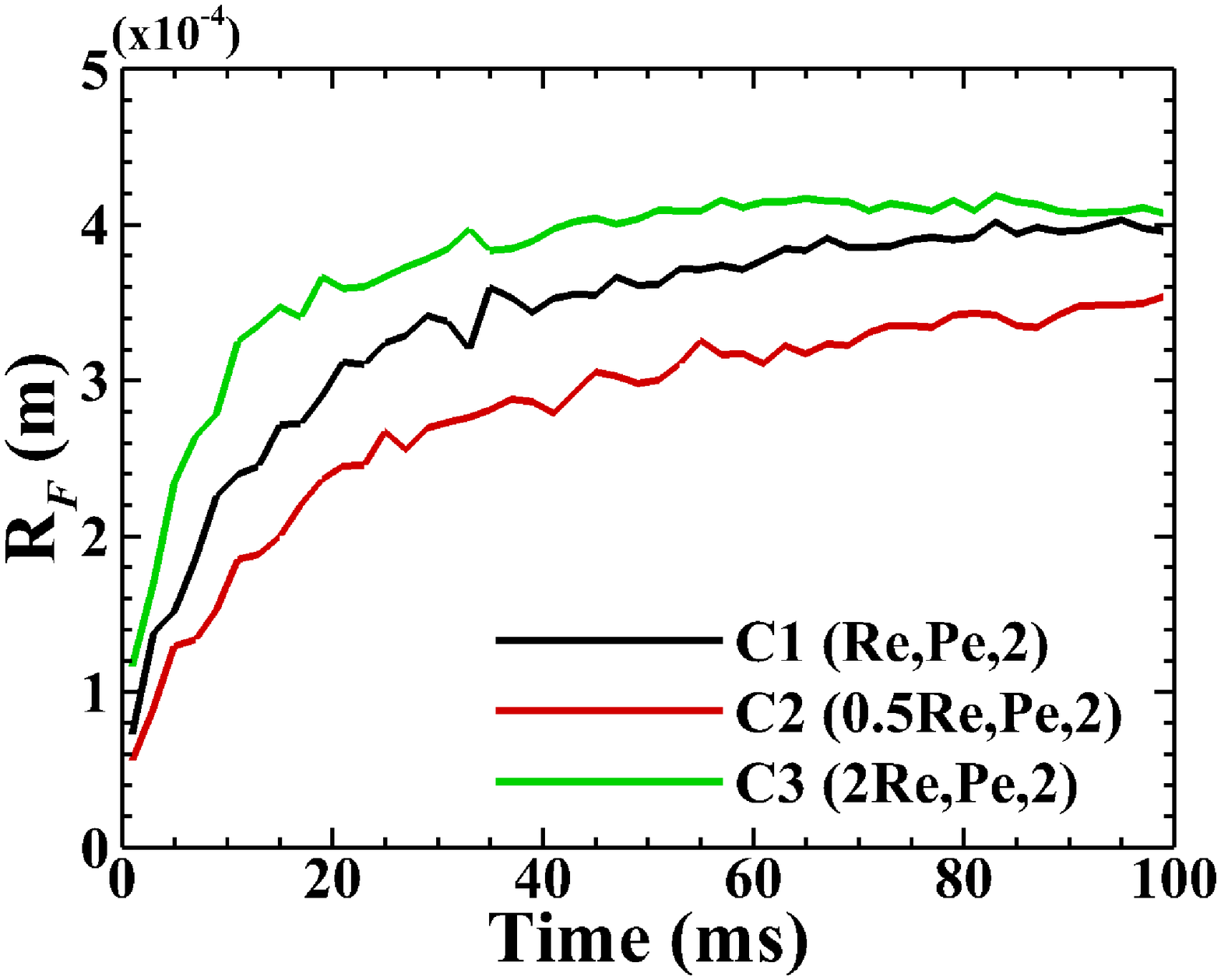}%
\caption{\label{fig:7} Finger growth versus time for different $Re$.}
\end{figure}

\begin{figure}  
\includegraphics[width=0.99\linewidth]{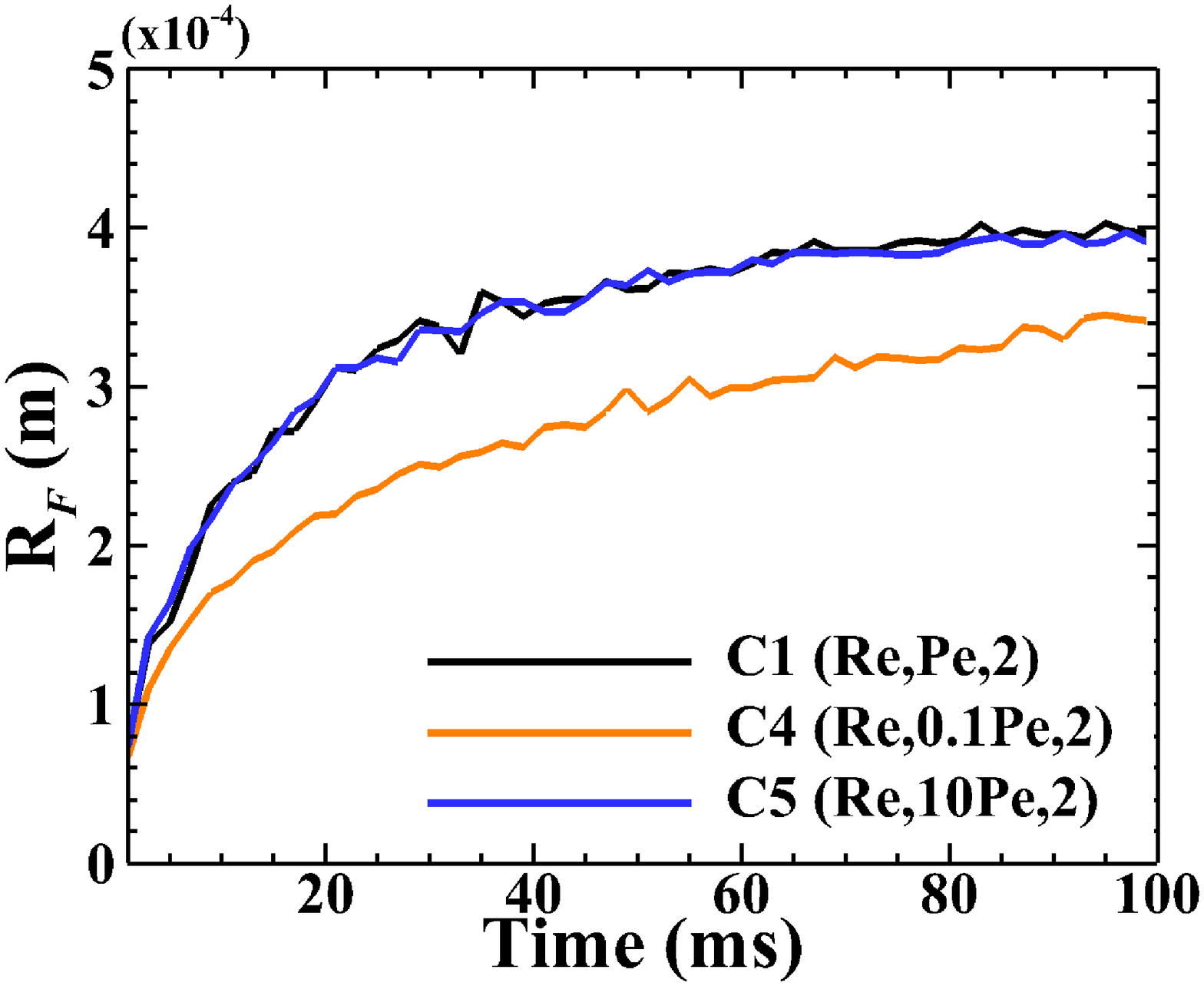}%
\caption{\label{fig:8} Finger growth versus time for different $Pe$.}
\end{figure}

\begin{figure}  
\includegraphics[width=0.99\linewidth]{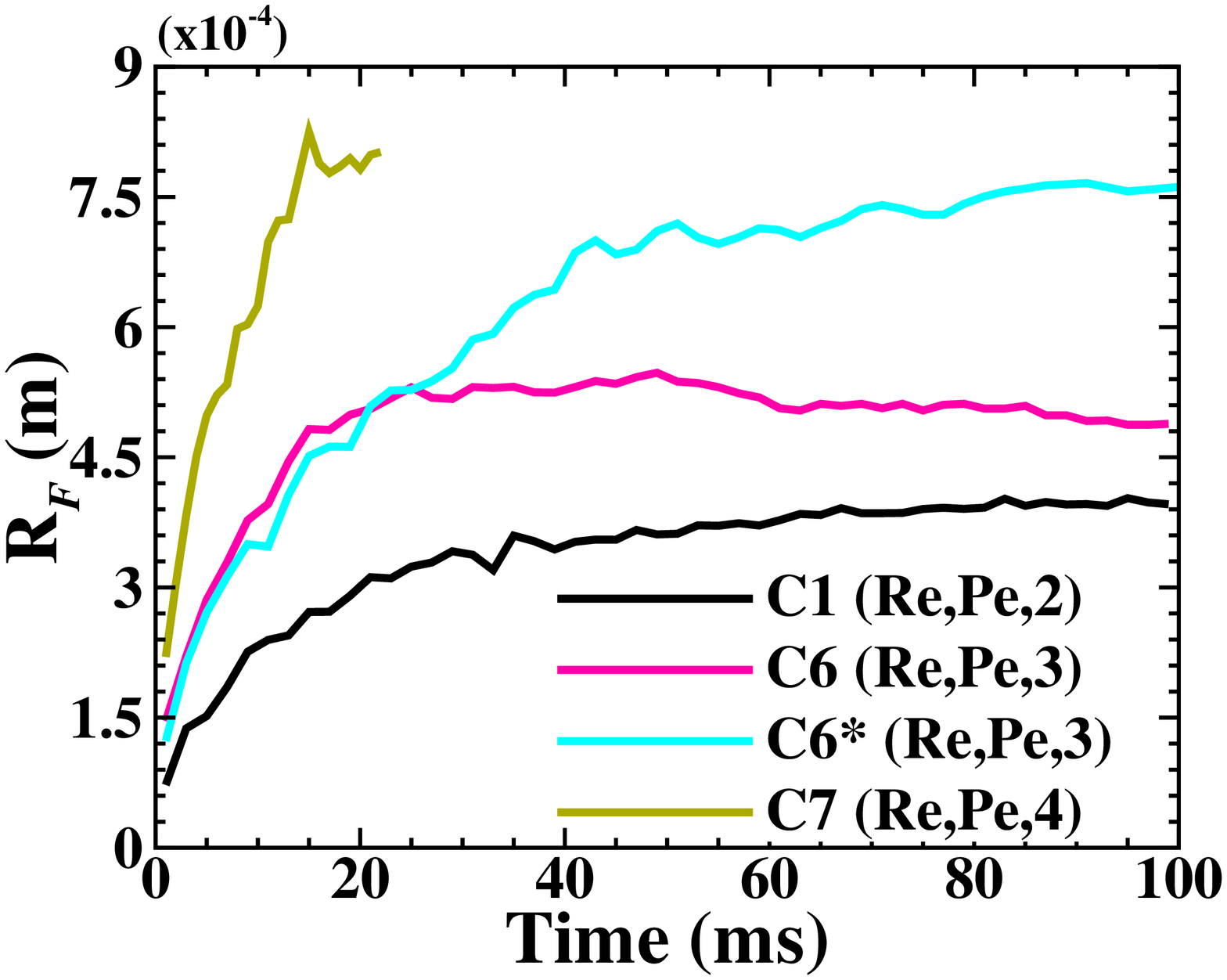}
\caption{\label{fig:9} Finger growth versus time for different of $\mathfrak{R}$.}
\end{figure}

For the quantitative analysis of the instabilities, we measure the extremities of the fingers by marking the circles ($R_I, R_O$) at the farthest and nearest point of fluid interaction (see figure \ref{fig:2}). Using these dimensions, we calculated the finger length ($R_F$=$R_O-R_I$) and the extent of the maximum fluid displaced from the domain ($R_I$). For the cases mentioned in table \ref{tab:table2}, we examine the degree of instability, sweep efficiency, and breakthrough characteristics. The fingering growth ($R_F$) over time is shown in different plots as a function of the non-dimensional parameters such as the Reynolds number (figure \ref{fig:7}), the P\'eclet number (figure \ref{fig:8}), and the mobility ratio (figure \ref{fig:9}).\\

As shown in figure \ref{fig:7}, an increase in the influx generates early and pronounced instability patterns that intensify with time(see case C3(2$Re, Pe$,2)). Fingers repeatedly split owing to local instability resulting in narrow fingers that move downstream with coalescence. The slow-growing fingers for lower influx (case C2) eventually sweep out the encapsulated fluid more efficiently. This finding concurs with the qualitative findings of \citep{chen1987radial} that a low flow rate displaces encapsulated fluid more efficiently. At a later time, the rate of growth of the fingers for cases C1, C2, and C3 becomes approximately similar.\\

The saturation iso-contour plots (figure \ref{fig:4}(a) and (b)) for Case C1, C4, and C5 at different times demonstrate that the instabilities have a noticeable dependency on the P\'eclet number. For diffusion coefficients of the order $10^{-8}$ or higher (where $Pe$ $\geq$ $O$(4)), the length of the finger remains the same (see figure \ref{fig:8}). Also, the finger growth increases monotonically for C4($Re,0.1Pe$,2), while for cases C1($Re, Pe$,2) and C5($Re,10Pe$,2), it rapidly increases initially but becomes uniform eventually. We find an early onset of finger patterns for flow with higher $Pe$ (case C1 and C5). The evolution of these cases over time shows extensive tip splitting and coalescence before the breakthrough. However, after coalescence, instabilities' expansion is delayed by shielding and spreading. As a result, the flow becomes less chaotic. In contrast, C4 has fewer fingers owing to higher diffusion. It is also reported \citep{pramanik2015effect} that the instability is more prominent with larger $Pe$ due to slow fluid transport for rectilinear Hele-Shaw cells. Additionally, linear stability analysis is used by \citep{tan1987stability} to explain the stability criterion based on $Pe_c$ (critical P\'eclet number) for radial source flow and concluded that the flow becomes unstable when $Pe$ $>$ $Pe_c$. Since we found that the flow is stable for C4 (lower Pe than other cases), whereas C1 and C5 exhibit instabilities, it indicates that for the range of our considered $Pe$ values, there will be some critical value of $Pe$ at which the stability behavior will change. Despite having an order of magnitude difference in $Pe$, cases C1 and C5 demonstrate similarity in finger growth. This observation signifies that the variation in the fingering instability is negligible beyond a certain value of the diffusion coefficient provided the $Re$ and the logarithmic viscosity ratio remain the same. \cite{suekane2017three} also reported similar unchanged fingertip extensions for a fixed viscosity ratio and an increasing P\'eclet number.\\
 
\begin{figure}  
\includegraphics[width=0.99\linewidth]{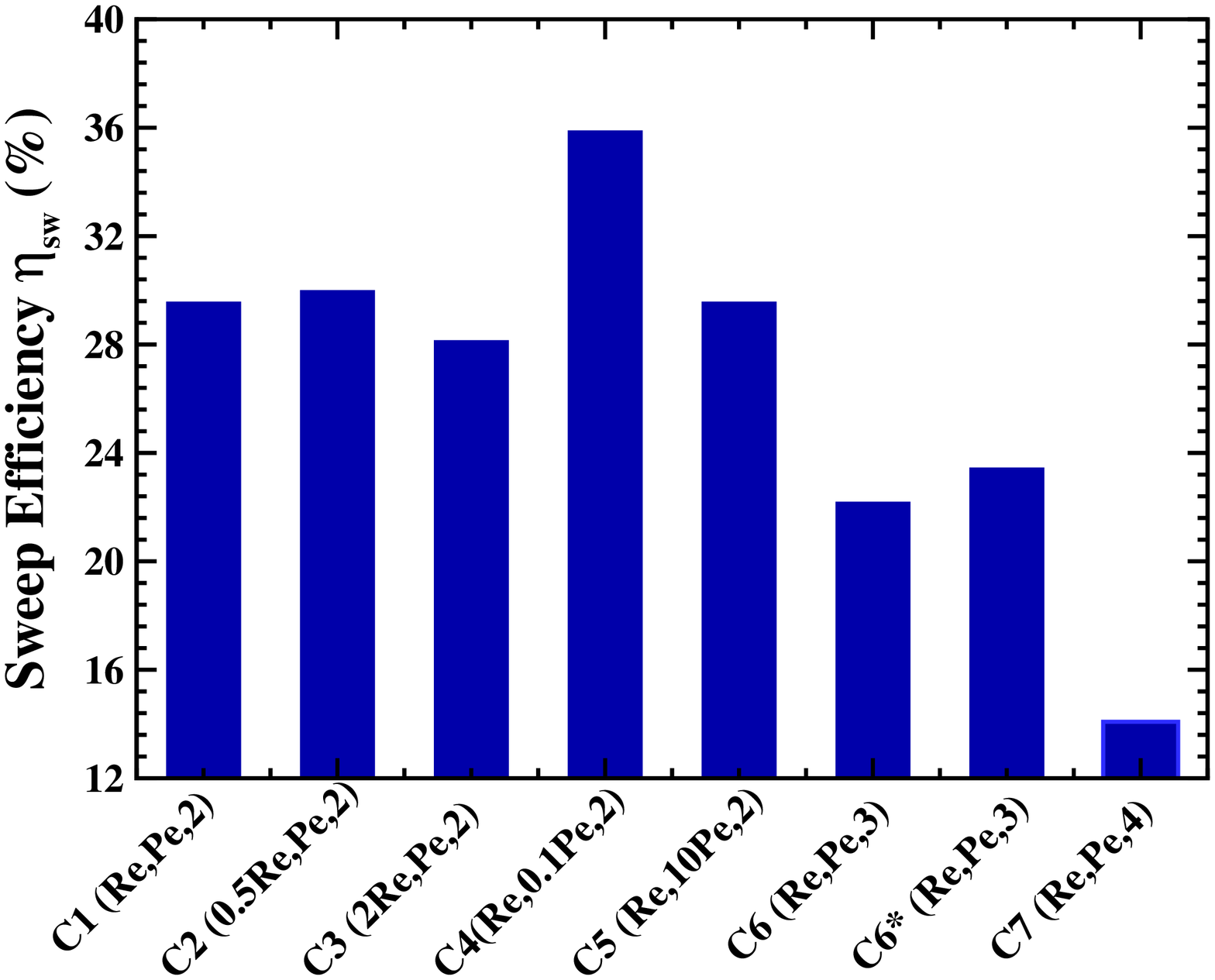}
\caption{\label{fig:11}Comparison of Sweep Efficiency ($\eta_{sw}$) for different cases}
\end{figure}

To maximize the sweep efficiency, researchers have explored the effect of different LVF to HVF ratios \citep{suekane2017three,bischofberger2014fingering}. By adjusting the non-dimensional number $\mathfrak{R}$ in figure \ref{fig:9}, we assess the impact of this variation on finger length. The initial growth rates for cases C1 and C6 are similar, but for case C7, a chaotic growth of the fingers is initially observed. Subsequently, due to the erratic expansion of the fingers for high viscosity difference in LVF and HVF (case C6 and C7), there is a drastic increment in the finger length. However, for case C6 the growth of the fingers ceases later, probably owing to the smaller size of the domain. To ensure that the growth of the fingers is indeed constrained by the size of the domain for C6, another case (C6$^*$) is simulated with the same flow parameters and properties as in C6, except that the size of the cubical domain is twice that of C6. The rate of growth of the fingers initially increases and becomes asymptotic at later times for C$6^*$. Surprisingly, we found that the size of the domain barely affects the sweep efficiency ($\eta_{sw}$) that remains similar for both C6 and C6$^*$ (see figure \ref{fig:11}). For the rest of the cases, the finger length keeps increasing (figures \ref{fig:7}, \ref{fig:8}, \ref{fig:9}), signifying that the domain effects are negligible on the growth of the fingers. The growth of fingers for C7 is highly non-linear and requires a significant amount of computational time. Therefore, we show the finger growth for C7 in figure \ref{fig:9} till the first breakthrough.\\


\begin{figure}  
\includegraphics[width=0.99\linewidth]{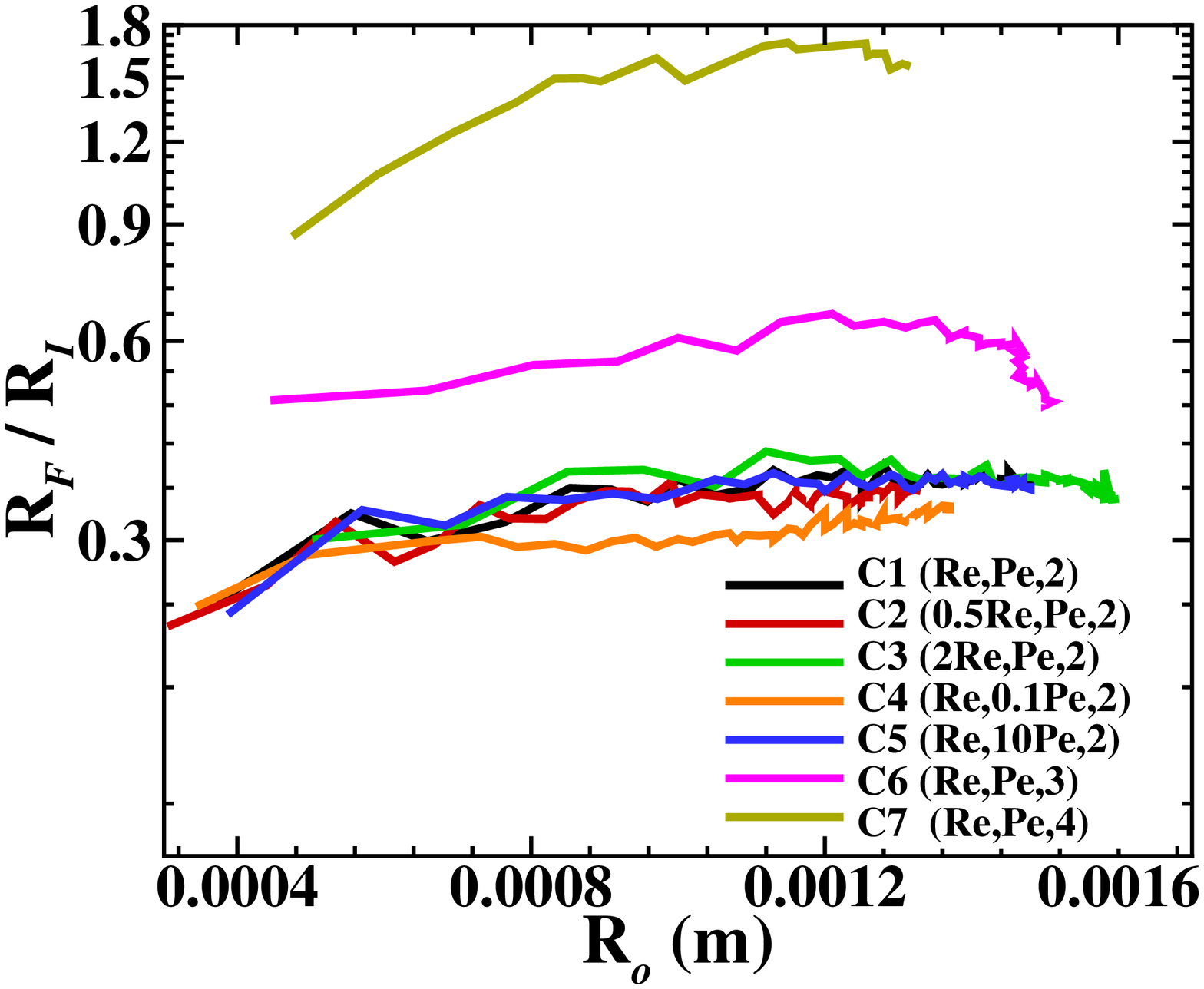}
\caption{\label{fig:10}Comparison of the extent of finger growth to the maximum total sweep-out fluid ($R_F/R_I$) versus  most distantly displaced fluid $R_O$.}
\end{figure}

These fingering patterns grow with time and move downstream by making channels without extracting the complete encapsulated HVF. $R_I$ (figure \ref{fig:2}) signifies the complete extraction of fluid. Enhanced oil recovery (EOR) techniques aim to maximize the sweep effectiveness (increasing $R_I$)\citep{kargozarfard2019viscous},\citep{riaz2003three},\citep{suekane2017three},\citep{suekane2019three}. Figure \ref{fig:10} compares the extent of finger growth to the maximum total sweep-out of HVF ($R_F/R_I$) to most distantly displaced fluid ($R_O$). It helps to explore the ideal combination of all the parameters that may result in optimal HVF extraction. For a given value of $R_O$, C7($Re, Pe$,4) demonstrates the highest value of $R_F$/$R_I$, indicating the least sweeping of HVF. The lower value of $R_F$/$R_I$ signifies a lesser number of instabilities and a higher value of $\eta_{sw}$ (see figure \ref{fig:11}). The flow is less stable if the LVF injection rate is high or the diffusion coefficient is low (see cases C3 and C5). In all the cases, the slope of the $R_F$/$R_I$ is initially high, indicating that at that instant, the finger length is more than the complete sweeping($R_I$). Coalescence of the fingers makes the fingers shorter and enhances the spreading of LVF in the domain.\\

\begin{table}[]
  \caption{\label{tab:table3}
First breakthrough time(ms) for different cases}
\begin{ruledtabular}
\begin{tabular}{cc}
\textbf {Cases}& ms \\\\
\hline
C1($Re,Pe$,2) & 44.5\\
C2($0.5Re,Pe$,2) & 81.59\\
C3($2Re,Pe,2$) &25.14 \\
C4($Re,0.1Pe,2$) & 52.16\\
C5($Re,10Pe,2$) &44.5 \\
C6($Re,Pe$,3) & 32.9\\
C7($Re,Pe$,4) & 22.2\\
\end{tabular}
\end{ruledtabular}
\end{table}

Researchers have made several technological advances to smoothly and quickly sweep HVF to achieve breakthrough \citep{garaeva2019numerical}, \citep{sajadi2021simulation}, \citep{gogoi2019review}, \citep{morrow2019numerical}. The instability patterns depend on tip splitting, coalescence, displacing fluid injection rate, and the dominant force (viscous or diffusive force). These factors also influence the downstream flow and hence the breakthrough characteristics. The first breakthrough time for all the cases is given in table \ref{tab:table3}. Early breakthrough can occur due to highly chaotic patterns, slow diffusion of fluid, or high LVF injection rate. Case C2 has a low-pressure gradient and a slow rate of injection of LVF, while Case C4 has better mixing and a higher diffusion rate, which means that the flow will take longer to reach downstream. For C3, breakthrough occurs comparatively early due to the rapid movement of the high-velocity LVF, whereas for C6 and C7, chaotic finger growth reaches downstream relatively quickly. A balance between breakthrough and sweep characteristics is required to achieve optimal HVF sweeping. Therefore, we have calculated the sweep efficiency ($\eta_{sw}$) for all the cases at the time of their first breakthrough. We define the sweep efficiency ($\eta_{sw}$) as the ratio of the volume of the LVF injected at the time of breakthrough to the volume of the cubical domain. To evaluate the volume of the injected LVF, we take the interface saturation value to $0.15$ \citep{suekane2017three}. Figure \ref{fig:11} demonstrates the sweep efficiency ($\eta_{sw}$) for all the cases at their first breakthrough time. The case with a higher diffusion coefficient (Case C4) or slow sweeping (Case C2) takes more time to sweep out the encapsulated HVF. The flow is stable for these cases and therefore has comparatively higher $\eta_{sw}$. The $\eta_{sw}$ of C4 with $Pe$ = 1250 and $\mathfrak{R} = 2$ is comparable with that reported by \cite{riaz2003three} for $Pe$=800 and $\mathfrak{R} = 2$. Additionally, we find a reduction in $\eta_{sw}$ with increasing $Pe$ similar to that demonstrated by \cite{suekane2017three, pramanik2015effect}. Moreover, C1 and C5 have similar values of $\eta_{sw}$, indicating that the effect of the variation in $Pe$ after a certain value is negligible \citep{tan1987stability}. The logarithmic mobility ratio, $\mathfrak{R}$, is an important parameter for $\eta_{sw}$, as with a larger difference in viscosity, it drastically decreases (Case C1 to C6 and C7). Therefore, for optimum oil extraction or similar applications, tuning all these parameters is imperative.\\
 
\section{\label{sec:conclusion}Conclusions}
We perform numerical simulations to study the influence of different parameters such as the $Re$, the $Pe$, and the logarithmic mobility ratio, $\mathfrak{R}$, on the dynamics of viscous fingering in a three-dimensional cubical domain. Additionally, we assess the role of the fingers in sweeping the encapsulated high-viscosity fluid out of the domain. This investigation has applications in the oil industry, where oil is extractde by injecting a low-viscosity fluid. We inject a low-viscosity fluid from a cylindrical hole into a cubical domain with a porous media. The $Re$ associated with this problem is defined based on the diameter of this hole, the injecting velocity, and the viscosity of LVF. Similarly, we define $Pe$ as the rate of advection to the rate of diffusion of the LVF to the HVF. We present a qualitative and quantitative assessment of the growth of the fingering instability in terms of the iso-surface contours and the extremities of the fingers. At high $Re$, the fingers evolve non-uniformly, resulting in earlier tip splitting and breakthrough compared to low $Re$. Owing to the uniform growth of the fingers at low $Re$, the LVF displaces the encapsulated HVF more efficiently. Similarly, for lower values of the $Pe$, the diffusion of LVF is high resulting in the ceasing of the fingering pattern and an efficient sweeping of HVF compared to the cases with larger $Pe$. We also observe that for lower values of $\mathfrak{R}$, the fluid expands uniformly and sweeps HVF more efficiently, as compared to higher values of $\mathfrak{R}$, which causes chaotic and non-linear growth of fingers. We further evaluate and report the sweep efficiency of all the cases. We observe that the sweep efficiency is high for the cases with a low injection velocity or high diffusion coefficient. This finding is attributed to a stable flow with fewer fingers developed within the domain. In contrast, the sweep efficiency drastically reduces with an increase in the ratio of the viscosity of the displaced to the displacing fluid owing to the development of finer fingers that interact and grow non-linearly. Therefore, to achieve an efficient sweeping of the high-viscosity fluid out of the domain, all the associated parameters should be combined such that the non-linear growth of the fingers is prohibited. \\

\nocite{*}

\bibliography{references}

\end{document}